\documentclass[useAMS,usegraphicx, usenatbib]{mn2e}
\usepackage{times}
\usepackage{aas_macros}

\usepackage{amsmath}
\usepackage{amssymb}
\usepackage[utf8x]{inputenc}
\usepackage{graphicx}
\usepackage{epsfig}
\usepackage{latexsym}
\usepackage{color}
\usepackage{dcolumn}
\usepackage{bm}
\usepackage{natbib}
\usepackage{setspace}
\usepackage{subfigure}
\usepackage{psfrag}
\usepackage{listings}
\usepackage{mathrsfs}
\usepackage{enumitem}

\definecolor{Red}{rgb}{1, 0, 0}
\definecolor{Green}{rgb}{0, 1, 0}
\definecolor{Blue}{rgb}{0, 0, 1}
\definecolor{Black}{rgb}{0, 0, 0}
\definecolor{Grey}{rgb}{0.5, 0.5, 0.5}
\definecolor{White}{rgb}{1, 1, 1}
\definecolor{Yellow}{rgb}{1, 1, 0}
\definecolor{Magenta}{rgb}{1, 0, 1}
\definecolor{Cyan}{rgb}{0, 1, 1}
\definecolor{myCyan}{rgb}{0, 0.6, 1}
\definecolor{Orange}{rgb}{1, 0.5, 0}
\definecolor{Violet}{rgb}{0.5, 0, 0.5}
\definecolor{DarkRed}{rgb}{0.5, 0., 0.3}
\definecolor{Pink}{rgb}{1., 0.4, 0.7}
\definecolor{LightPink}{rgb}{1, 0.8, 0.8}
\definecolor{YellowGreen}{rgb}{0.6, 0.8, 0}
\definecolor{LightYellow}{rgb}{1., 1., 0.95}
\definecolor{Brown}{cmyk}{0, 0.8, 1, 0.6}

\begin{document}
\pagestyle{myheadings}
\markboth{Gas properties in dark matter mini-haloes}{}

 \title{Robust PCA and MIC statistics of baryons in early mini-haloes}

\author[R.~S.~de Souza, U.~Maio, V.~Biffi, B.~Ciardi]
{R. S. de Souza$^{1,2}\thanks{e-mail: rafael.2706@gmail.com}$,   U. Maio$^{3,4}$, V. Biffi$^{5}$,  B. Ciardi$^{6}$
\\
$^{1}$Korea Astronomy \& Space Science Institute, Daedeokdae-ro 776,  305-348  Daejeon, Korea\\
$^{2}$MTA E\"otv\"os University,  EIRSA "Lendulet" Astrophysics Research Group, Budapest 1117, Hungary\\
$^{3}$INAF - Osservatorio Astronomico di Trieste, Villa Bazzoni via G. B. Tiepolo 11, I-34143 Trieste, Italy\\
$^{4}$Leibniz Institute for Astrophysics, An der Sternwarte 16, D-14482 Potsdam, Germany\\
$^{5}$SISSA - Scuola Internazionale Superiore di Studi Avanzati, Via Bonomea 265, 34136 Trieste, Italy\\
$^{6}$Max-Planck-Institut f\"ur Astrophysik, Karl-Schwarzschild-Str. 1, D-85748 Garching, Germany\\
}


\pagerange{\pageref{firstpage}--\pageref{lastpage}} \pubyear{2013}

\maketitle
\label{firstpage}

\begin{abstract}

We present a novel approach, based on  robust principal components analysis (RPCA) and maximal information coefficient (MIC), to study the redshift dependence of halo baryonic properties. Our  data are composed of a set of different physical quantities for primordial minihaloes: dark-matter mass ($M_{\mathrm{dm}}$), gas mass ($M_{\mathrm{gas}}$), stellar mass ($M_{\mathrm{star}}$), molecular fraction (${\mathrm{x_{mol}}}$), metallicity (\textit{Z}), star formation rate (SFR) and temperature.
We find that $M_{\mathrm{dm}}$ and $M_{\mathrm{gas}}$ are dominant factors for variance, particularly  at high redshift. Nonetheless, with the emergence of the first stars and subsequent feedback mechanisms, ${\mathrm{x_{mol}}}$, SFR and \textit{Z} start to have a more dominant role.  Standard PCA gives three principal components (PCs)  capable to explain more than 97 per cent of the data variance at any redshift  (two PCs usually accounting  for no less than  92 per cent), whilst  the first PC  from the  RPCA analysis  explains no less than 84  per cent of the total variance in the entire redshift range (with two PCs explaining  $ \gtrsim 95$ per cent anytime). 
Our  analysis also  suggests that all the gaseous properties have a stronger correlation with $M_{\mathrm{gas}}$ than with $M_{\mathrm{dm}}$, while $M_{\mathrm{gas}}$ has a deeper correlation with ${\mathrm{x_{mol}}}$ than with \textit{Z} or SFR.
This indicates the crucial role of gas molecular content to initiate star formation and consequent metal pollution from Population III and Population II/I regimes in primordial galaxies.
Finally, a comparison between MIC and Spearman correlation coefficient shows that the former is a more reliable indicator when halo properties are weakly correlated.
\end{abstract}


\section{Introduction}
The standard model of cosmology predicts a structure 
formation scenario driven by cold dark matter \citep[e.g.,][]{Benson2010}, where galaxies form from molecular gas cooling within growing dark matter haloes.
Hence, understanding the correlation between different properties of the dark matter haloes is imperative to build up a comprehensive  picture of galaxy evolution.
Many authors have explored the correlation between dark-halo properties, such as mass, spin and shape, both in low- \citep[e.g.,][]{Bett2007,Hahn2007, Maccio2007,Wang2011} and high-redshift \citep[e.g.,][]{Hannah2001,deSouza2013a} regimes.
Estimating the strength of these correlations is critical to support  semi-analytical and halo occupation models, which assume the mass as determinant factor of the halo properties
\citep[e.g.,][]{Mo1996,Cooray2002,Berlind2003,Somerville2008}.
Nevertheless, alternative approaches, based on principal components analysis (PCA), found that concentration is a key parameter, contrary to what expected before \citep{Jeeson2011,Skibba2011}, and stressed the need for further investigations.
PCA belongs to a family of techniques ideal to explore high-dimensional data. The method consists in projecting the data into a low-dimensional form, retaining as much information as possible \citep[e.g.,][]{Jollife2002}.
Hence, PCA emerges as a natural technique to investigate correlation and temporal evolution of halo properties.
Because of  its versatility, PCA has been applied to a broad range of astronomical studies, such as 
stellar, galaxy and quasar spectra \citep[e.g.,][]{Chen2009, McGurk2010}, 
galaxy properties \citep{Conselice2006, Scarlata2007}, 
Hubble parameter and cosmic star formation (SF) reconstruction \citep[e.g.,][]{ishida2011a, ishida2011b}, and
supernova (SN) photometric classification \citep{ishida2013}.

Despite its generality,  PCA is not the only way to handle huge  data sets,  and
the growth in complexity of scientific experimental data makes the ability to extract newsworthy and meaningful information an endeavor per se.
The yearning for novel methodologies of data-intensive science gave rise to the so-called fourth research paradigm \citep[e.g.,][]{Bell2009}.
Data mining methods have been used in many areas of knowledge such as genetics \citep[e.g., ][]{Venter2004} and financial marketing decisions \citep[e.g., ][]{Shaw2001}, and their importance for astronomy has been recently highlighted as well \citep[e.g.,][]{Ball2010, Graham2013, KISS2013,Martinez2013}.
Likewise observations, cosmological simulations are continuously increasing in complexity, lessening the distance between observed and synthetic data \citep[e.g., ][]{Overzier2013, deSouza2013b,deSouza2014}.
None the less, the application of data-mining to cosmological simulations remains a \textit{terra incognita}.

In this work, we investigate the statistical properties of baryons inside high-redshift  haloes, including detailed chemistry, gas physics and stellar feedback. We make use of Robust PCA (RPCA) and maximal information coefficient (MIC) to study a set of various halo parameters.  RPCA represents a generalization of the standard PCA, whose advantage is its resilience to outliers and skewed data, while MIC is expected to be the correlation analysis of the 21st century \citep{Speed2011}, in particular due to MIC ability in quantifying general associations between variables.
Therefore, this project represents the first application of MIC to $N$-body/hydro simulations, and the first use of PCA to explore the low-mass end of the halo mass function and the birth of the first galaxies.

The outline of this paper is as follows.
In Section \ref{sec_sim}, we describe the cosmological simulations and their  outcomes.
In Section \ref{sec_statistics}, we describe  the statistical methods.
In Section \ref{sec_results}, we present our analysis and main results.
Finally, in Section \ref{sec_conclusions}, we present our conclusions.

\section{Simulations}
\label{sec_sim}

We analyzed the results of a cosmological $N$-body, hydrodynamical, chemistry simulation based on \citealt{Biffi2013} (see also  \citealt{maio2010, maio2011a}), that was run by means of a modified version of the smoothed-particle hydrodynamics code \textsc{gadget2} \citep{Springel2005}.
The modifications include relevant chemical network to self-consistently follow the evolution of e$^-$, H, H$^+$, H$^-$, He, He$^{+}$, He$^{++}$, H$_2$, H$_2^+$, D, D$^+$, HD, HeH$^+$ 
\cite[e.g.,][]{yoshida2003, maio2006, maio2007, maio2009},
ultraviolet background radiation, metal pollution according to proper stellar yields (He, C, O, Si, Fe, Mg, S, etc.), lifetimes and stellar population for Population III (Pop III) and Population  II/I (Pop II/I) regimes \cite[][]{tornatore2007}, radiative gas cooling from molecular, resonant and fine-structure transitions \cite[e.g.][ and references therein]{maio2007} and stellar feedback  \cite[][]{springel2003}.
The transition from the Pop III to the Pop II/I regime is determined by the value of the gas metallicity ($Z$) compared to the critical value $Z_{crit}$  \citep[e.g., ][]{Omukai2000, Bromm2001},  assumed to be $10^{-4}Z_{\bigodot}$\footnote{
Although uncertain \citep{bromm2003, Schneider2003, Schneider2006}, results are usually not very sensitive to the precise value adopted \cite[][]{maio2010}.}.

The cosmic field is sampled at redshift $ z = 100 $, adopting standard  cosmological parameters:  $\Omega_{\Lambda} = 0.7, \Omega_{m} = 0.3, \Omega_{b} = 0.04, H_{0}= 70$ km/s/Mpc and $\sigma_8 =  0.9$.
We considered snapshots in the range $9 \lesssim z \lesssim 19$, within a cubic volume of comoving side 0.7  Mpc, and $2 \times 320^{3} $ particles per gas and dark-matter species corresponding to particle masses of  42  and $\rm 275~M_{\bigodot} {\it h^{-1}}$, respectively.
The identification of the simulated objects is done by applying a friends-of-friends (FoF) technique with linking
length equal to 20 per cent the mean interparticle separation and substructures are identified by using a {\sc subfind} algorithm \cite[][]{Dolag2009}, which discriminates among bound and non-bound particles. The halo characteristics, such as position, velocity, dark matter and baryonic properties,  are computed and stored at each redshift.

The simulation outcomes investigated here consist of seven parameters: dark-matter mass ($M_{\mathrm{dm}}$), gas mass ($M_{\mathrm{gas}}$),  stellar mass ($M_{\mathrm{star}}$), star formation rate (SFR), $Z$, gas temperature (\textit{T}), and gas molecular fraction $(\mathrm{x_{mol}})$.
We refer the reader to   previous works, where more details and additional analyses about
halo spin and shape distribution \cite[][]{deSouza2013a},
feedback mechanisms \cite[][]{maio2011a, pm2012, maio2013},
primordial streaming motions \cite[][]{maio2011c},
non-standard cosmologies \cite[][]{maio2006, maio2011b, maio2011d, deSouza2013c},
high-$z$ luminosity function \cite[][]{salvaterra2013, Dayal2013tmp},
early gamma ray bursts- \cite[][]{campisi2011,deSouza2011a, deSouza2012, maio2012,Mesler2014}  and SNe-host properties \cite[][]{deSouza2010,deSouza2011b,Johnson2013,Whalen2013a,Whalen2013b},
Ly$\alpha$ emitters \cite[][]{akila2012}
and damped Ly$\alpha$ (DLA) system chemical content \cite[][]{maio2013arXiv}
are presented and discussed.

\subsection{Data set}

\begin{figure}
\centering
\includegraphics[width=1\columnwidth]{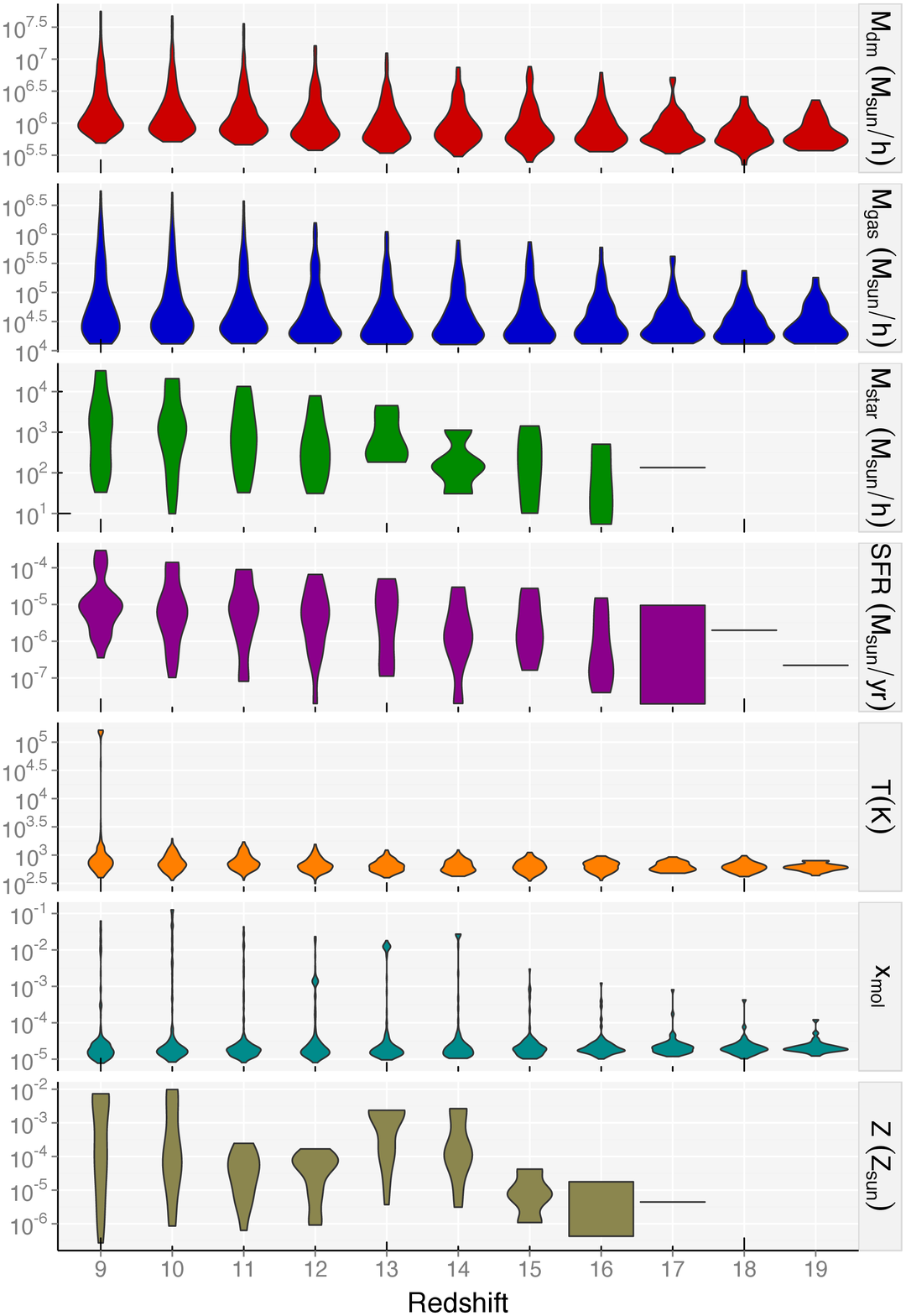}
\caption{Redshift evolution of halo properties.  From top to bottom: $\rm M_{dm}$ (red), $\rm M_{gas}$ (blue), $\rm  M_{star}$ (green), \textit{SFR} (magenta), T (orange), $\mathrm{ x_{mol}}$ (cyan) and Z (khaki).  They are  portrayed by a  violin
plot.  Each violin centre represents the median of the distribution, 
while the shape,   its mirrored PDF. 
 }
\label{fig:variables_evol}
\end{figure}

The total dataset   is composed by a few thousands haloes 
at very high redshift,  $z \approx 19$, and reaches about   25000 primordial objects
at $z \approx 9$.  In order to avoid numerical artifacts, created by a poor number of gas particles \citep{Bate1997}, we selected only those structures in which the gas content is resolved with at least $300$ gas particles. This usually corresponds to selecting only objects with a total number of particles of at least $\sim 10^3$. 
 The remaining  data  are therefore composed of  $\approx$1680  haloes in the whole   redshift  range, of which $\approx 200$ are   at  $z = 9$.
 Fig. \ref{fig:variables_evol} shows the  probability distribution function (PDF)
  for the  seven halo parameters: $M_{\rm dm}, M_{\rm gas}, M_{\rm star}$,  SFR, $T$, $\mathrm{x_{mol}}$ and  $Z$   at each redshift. They are  portrayed by a  violin
plot.  Each violin centre represents the median of the distribution, 
while the shape,   its mirrored PDF.  A visual inspection in Fig. \ref{fig:variables_evol}   indicates the first stages of  significant SF activity around $z = 17$,  giving  rise to a subsequent  boost in metal enrichment at $z \gtrsim 15$, and  a similar  growth of   $M_{\rm star}$ in the same redshift range.  Just after this episode, we can see the rapid spread in the $\mathrm{x_{mol}}$ variance, peaking few orders of magnitude above average.  The masses of the  haloes  range between     $ 10^{5} M_{\bigodot} \lesssim M_{\rm dm} \lesssim 10^{8} M_{\bigodot}$ and  $ 10^{4} M_{\bigodot} \lesssim M_{\rm gas} \lesssim 10^{7} M_{\bigodot}$.
Typical temperatures range from  500 to  $10^4$ K, where $H_2$ shapes
the thermal conditions of early objects. Hotter temperatures
are due to the thermal effects of SN explosions that heat and enrich  the gas in nearby smaller haloes.

\section{Statistical Analysis}
\label{sec_statistics}

\subsection{Robust Principal Components Analysis.}

The ultimate goal of PCA is to reduce the dimensionality of a multivariate data\footnote{
  A  set of measurements on each of two or more variables.
}, while explaining the data variance  with as few principal components (PCs) as possible.
PCA belongs to a class of Projection-Pursuit \citep[PP; e.g.,][]{Croux2007}  methods, whose aim is to detect structures in multidimensional data  by projecting them into a lower-dimensional subspace (LDS). The LDS is selected by maximizing a projection index (PI), where PI represents an \textit{interesting feature} in the data (trends, clusters, hyper-surfaces, anomalies, etc.).
The particular case where  variance ($S^2$) is taken  as a PI leads to the  classical version of PCA\footnote{
  The PCs are computed by diagonalization of the data covariance matrix ($\Sigma^2$),  with the resulting eigenvectors corresponding to PCs and the resulting eigenvalues to the variance {\it explained} by the PCs.\\
The eigenvector corresponding to the largest eigenvalue gives the direction of greatest variance (PC1), the second largest eigenvalue gives the direction of the next highest variance (PC2), and so on.
Since covariance matrices are symmetric positive semidefinite, the eigenbasis is orthonormal (spectral theorem).
}.

Given $n$ measurements  $x_1, \cdots, x_n$, all of them column vectors of dimension $\Gamma$, the first PC is obtained by finding a unit vector $\mathbf{a}$ which maximizes the variance of the data projected on it:
\begin{equation}
\mathbf{a_1} =\underset{||\mathbf{a}||=1}{\rm \arg\max}~ S^2 (\mathbf{a}^tx_1,\cdots,\mathbf{a}^tx_n), 
\label{eq:PC1}
\end{equation}
where $t$ is the transpose operation and   $\mathbf{a_1}$ is the direction of the first PC\footnote{
  $\underset{x}{\arg\max}~  f(x)$ is the set of values of $x$ for which the function $f(x)$ attains its largest value.
}.
Once we have computed the $(k-1)$th PC, the direction of the $k$th component, for $1 < k \leqslant \Gamma$, is given by 
\begin{equation}
\mathbf{\mathbf{a}_k} = \underset{||\mathbf{a}||=1,\mathbf{a}\bot \mathbf{a}_1,\cdots,\mathbf{a}\bot \mathbf{a}_{k-1}}{\rm \arg\max}S^2(\mathbf{a}^tx_1,\cdots,\mathbf{a}^tx_n), 
\end{equation}
where the condition of each PC to be orthogonal to all previous ones, ensures a new uncorrelated basis.
In spite of these attractive properties, PCA has some critical drawbacks as the  sensitivity to outliers  \citep[e.g.,][]{Hampel2005},  and inability to deal with missing data \citep[e.g.,][]{Huan2010}.
In order to overcome this limitation,  several  robust versions were created  based on the PP principle. Instead of taking the variance as a PI in equation (\ref{eq:PC1}), a robust\footnote{
 Robust statistics commonly use median and median absolute deviation, instead of mean and standard deviation, in order to be resistant  against outliers.
} 
measure of variance is taken. Hereafter, we will refer the standard variance as $S^2_{\rm sd}$ and robust variance as  $S^2_{\mathrm{MAD}}$. 
Two common measures of robust variance \citep{Hoaglin2000} are the median absolute deviation \citep[MAD; e.g., ][]{Howell2005}, 
\begin{equation}
\mathrm{MAD}(\kappa_1, \cdots,\kappa_n) = 1.48 \underset{j}{\mathrm{med}}|\kappa_j - \underset{i}{\mathrm{med}}\kappa_i|, 
\end{equation}
and the first quartile of the pairwise differences between all data points \citep[\textit{Q}; e.g.,][]{Rousseeuw1993}, 
\begin{equation}
Q(\kappa_1, \cdots,\kappa_n) = 2.22 \left\{|\kappa_i - \kappa_j|;1\leqslant i < j \leqslant n\right\}_{\binom{2}{n}/4},
\end{equation}
where $\{\kappa_1,\cdots,\kappa_n\}$ is a given univariate dataset and the square of MAD or \textit{Q} gives a robust  variance\footnote{When the PI is the standard variance, the first PC  is the eigenvector of the data covariance matrix  corresponding to the largest eigenvalue. But this does not hold for general choices of variance and approximative algorithms are necessary. }.  Hereafter all calculations of the PCs are performed using   the grid  search base algorithm   \citep{Croux2007} with  MAD, but   using \textit{Q}  has no influence on our results. Also note that  before applying  the PCA, we standardize the halo
properties by subtracting the mean and dividing by the standard
deviation. Therefore we are formally using the correlation matrix that can be seen as the covariance matrix of standardized variables.

\subsection{Maximal information coefficient.}

The maximal information-based non-parametric exploration (MINE) statistics represent a novel family of  techniques to identify and characterize general relationships in data sets \citep{Reshef2011}. MINE  introduce MIC as a new measure of dependence between two variables, which   possesses  two desired properties for data exploration: (i) generality, the ability to capture a broad range of associations and functional relationships\footnote{
  For comparison,  Pearson  coefficient  measures the linear correlation between two variables, while Spearman coefficient ($R_s$)  measures the strength of monotonicity  between paired data.};
(ii) equitability, the ability to  give similar scores to equally noisy relationships of different types\footnote{
  In benchmark  tests,  MIC equitability behaves better than  other  methods  such as e.g., mutual information estimation, distance correlation and $R_s$.  A lack of equitability introduces  a strong bias and entire classes of relationships may be  missed \citep{Reshef2013}.}.

MIC measures the strength of general associations, based  on the mutual information\footnote{Mutual
information measures the general  interdependence between two variables,  while the correlation function
measures the linear dependence between them \citep[e.g.,][]{Wentian1990}.} ($\mathrm{MI}$)   between  two random variables $A$ and $B$: \footnote{MIC tends to 1 for all never-constant noiseless functional relationships and to 0 for statistically independent variables.}

\begin{equation}
\mathrm{MI}(A, B) = \sum_{a \in A} \sum_{b \in B} p(a,b) \log \left( \frac{p(a,b)}{p(a)p(b)} \right), 
\end{equation}
\noindent
where $p(a)$ and $p(b)$ are the marginal  PDFs
of $A$ and $B$, and $p(a,b)$ is the joint PDF.\\
Consider D a finite set of ordered pairs, $\{(a_i, b_i), i = 1, \ldots, n\}$,  partitioned into a $x$-by-$y$ grid of variable size, $G$, such that there are  $x$-bins spanning $a$ and $y$-bins covering $b$, respectively.
\\
The PDF of a particular grid cell is  proportional to the number of data points  inside that cell.
We can  define a  characteristic matrix $M(D)$ of a set $D$ as
\begin{equation}
\label{eq:MD}
\mathrm{M(D)}_{x,y} =  \frac{\max(\mathrm{MI})}{\log \min \{x,y\}},
\end{equation}
\\
representing the highest normalized mutual informations of $D$.
The MIC of a set $D$ is then defined as
\begin{equation}
\mathrm{MIC(D)} = \max_{0 < xy < B(n)} \left\{\mathrm{M(D)}_{x,y} \right\}, 
\end{equation}
representing the maximum value of $M$ subject to $0 < xy <  B(n)$, where the function $B(n) \equiv n^{0.6}$ was empirically determined by \citealt{Reshef2011}\footnote{
  The  $0.6$ exponent value represents a compromise since high values of $ B(n) $ lead to non-zero scores even for random data, as each point gets its own cell, while low values only probe simple patterns.}.

\section{Results}
\label{sec_results}

Hereafter we discuss the relations  between  halo properties  and  their relative importance. Our matrix is composed by   1680 haloes,  spanning  the redshift range $9 \lesssim z \lesssim19$,  with  $\approx$ 200 (30) haloes at $z$ = 9 (19), each halo  containing at least $\sim 10^3$ particles.
Each row of the matrix represents a halo and each column represents one of the halo properties. PCA probes the entire matrix at once. 
 On the other hand, MIC is a pair-variable comparison, therefore  requiring   $N(N-1)/2$ operations, with  $N$ being  the number of halo properties.  It is worth to highlight  here that each approach has its own advantages and disadvantages. PCA is suitable for high-dimensional data, when a pair comparison becomes unfeasible, however the method only searches for linear relationships.  MIC, instead, finds general associations in data structures, but may be  impractical to deal with a large amount of parameters.

\subsection*{PCA}

\begin{figure}
\centering
\includegraphics[width=1\columnwidth]{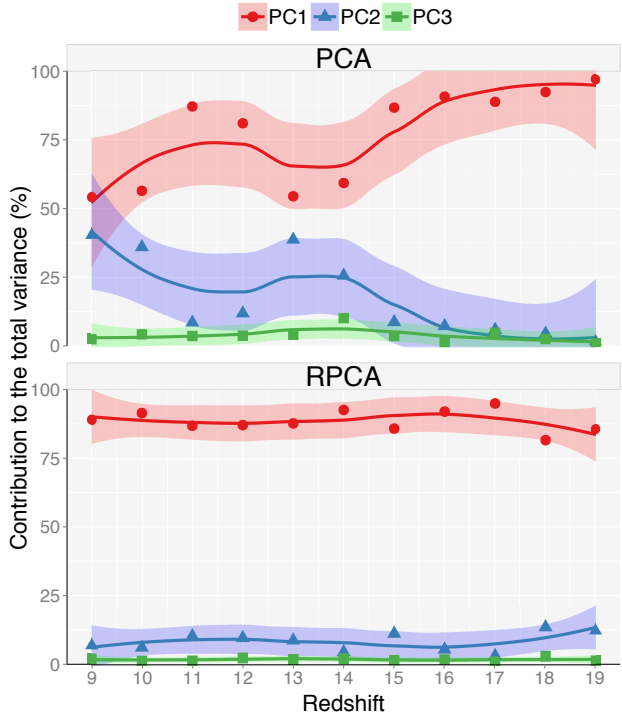}
\caption{Fraction of variance  explained by the first three PCs  as a function of redshift;  PC1 (red circles), PC2 (blue triangles), PC3 (green squares).  Symbols represent the actual estimate values for each snapshot, while the curves represent a smooth fitting with 95 per cent confidence level  limited by the shadowed areas. The curves and confidence levels are estimated by a local polynomial regression fitting \citep{Cleveland1992}. Top panel: PCA; bottom panel: RPCA. 
  }
\label{fig:PCA_evol}
\end{figure}

In order to better  understand the pros and cons of using  RPCA, we first start the analysis with the standard PCA. 
Fig.  \ref{fig:PCA_evol} shows the contribution of the first three PCs  to $S^2_{sd}$, as a function of redshift. Three  PCs account for more than $97$ per cent of $S^2_{sd}$ at any redshift, while two  PCs explain more than $92$ per cent except at $z \simeq 14$, when the contribution drops to $85$ per cent.

The sharp variation of the PCs around $z \simeq 14-16$ acts as a smoking gun for a global cosmological event.
Indeed, this is a direct consequence of first SF episodes and the interplay between chemical and mechanical feedback from the first stars, that takes place around $z\simeq 15-20$ \citep[e.g., ][]{maio2010, maio2011a}.
As molecules are produced over time, they lead to gas collapse, stellar formation and metal pollution, with consequent back reaction on the thermal behavior of the surrounding gas \citep[see e.g., ][]{maio2011a,Biffi2013}. 
This redshift range represents an epoch of fast and turbulent growth of the  metal filling factor, from $\sim 10^{-18}$ at $z \simeq 15$  to $\approx 10^{-12}$ at $z \simeq 14$ \citep[see Fig. 1 from][]{maio2011a}.
At the beginning, only the gas at high densities is affected by metal enrichment, due to  SF concentration in these regions. As SF and metal spreading proceed, the surrounding lower density environments are affected as well. SNe heat high-density gas within star-forming sites and, consequently, hot low-density gas is ejected from star-forming regions by SN winds.

The contribution of each PC  dramatically changes if we use RPCA instead. The clearest advantage is the amount of variance explained by each component (Hereafter, when necessary  to avoid ambiguity,  the  PCs from RPCA analysis will be referred as RPCs).   RPC1 accounts for  no less than $\approx 84$ per cent of the $S^2_{\mathrm{MAD}}$ anytime, whilst two RPCs account  for more than $\approx 95$ per cent. Moreover, the RPC2 contribution mostly stands out between at  $13 < z < 17$ and $ z \lesssim 10$. Albeit contributing differently to the total variance, the   general behavior of PC1  and PC2  is similar to the RPC1 and RPC2,  as well as  the physical interpretation. But RPCA assigns less weight to the baryonic properties, suggesting the halo mass as the most significant factor.  This difference occurs because   even a small
fraction of large errors can cause arbitrary corruption in PCA's estimate.  For instance, PCA   is more sensitive to rapid variations  of the   halo chemical  properties, having a steeper  reaction in their  first PCs.  Thus, as expected RPCA surpass PCA in their ultimate goal: reduce the system dimensionality.  Nevertheless, the greatest power to synthesize information carries the assumption that outliers are caused by  corrupted data,  which is not always the case. 
This potential  drawback will be better understood looking at the contribution of each variables to the $k$-th PCs  as discussed in the following.

\begin{figure*}
\begin{minipage}[b]{1\linewidth}
\centering
\includegraphics[width=0.48\columnwidth]{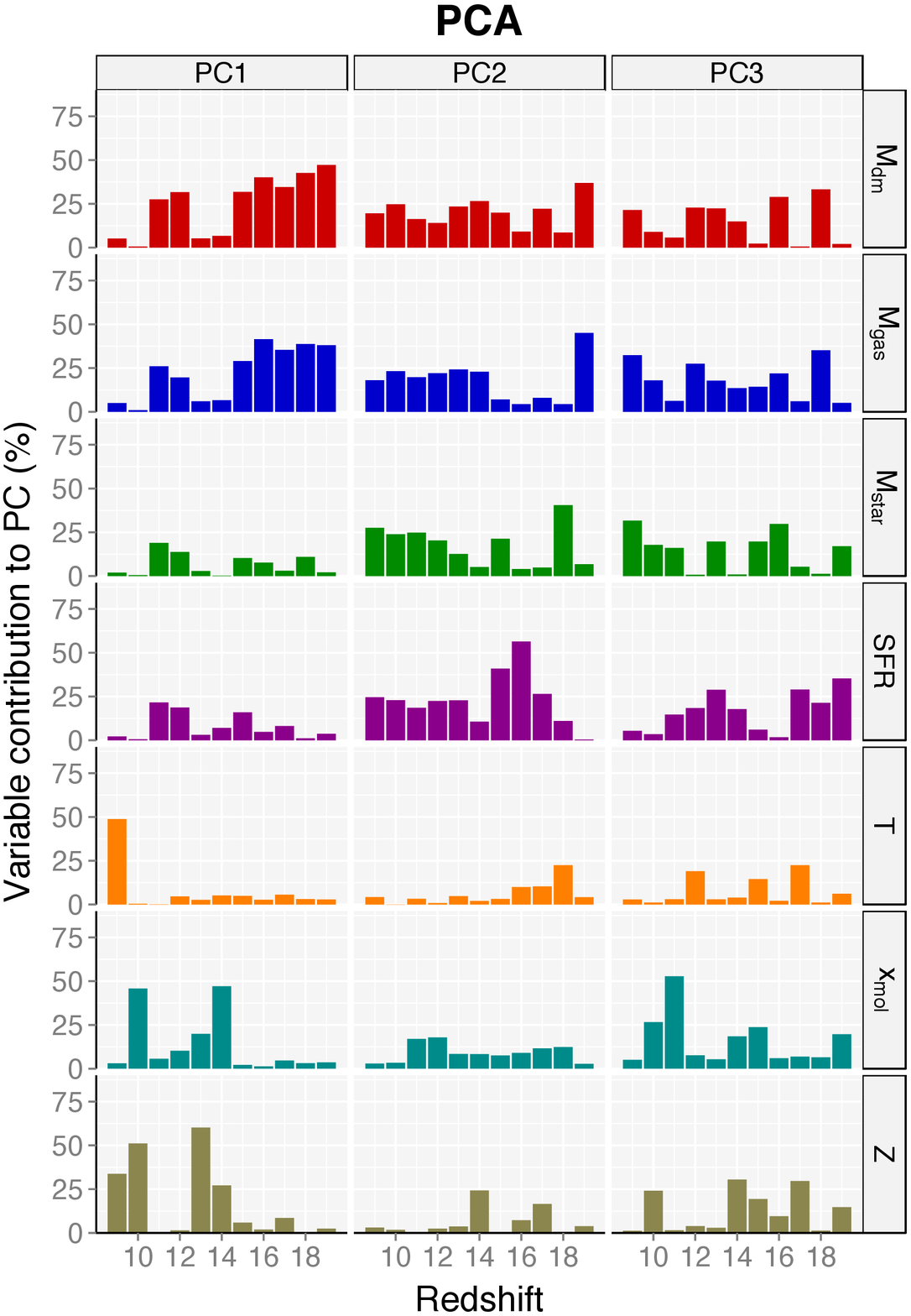}
\hspace{0.5 cm}
\includegraphics[width=0.48\columnwidth]{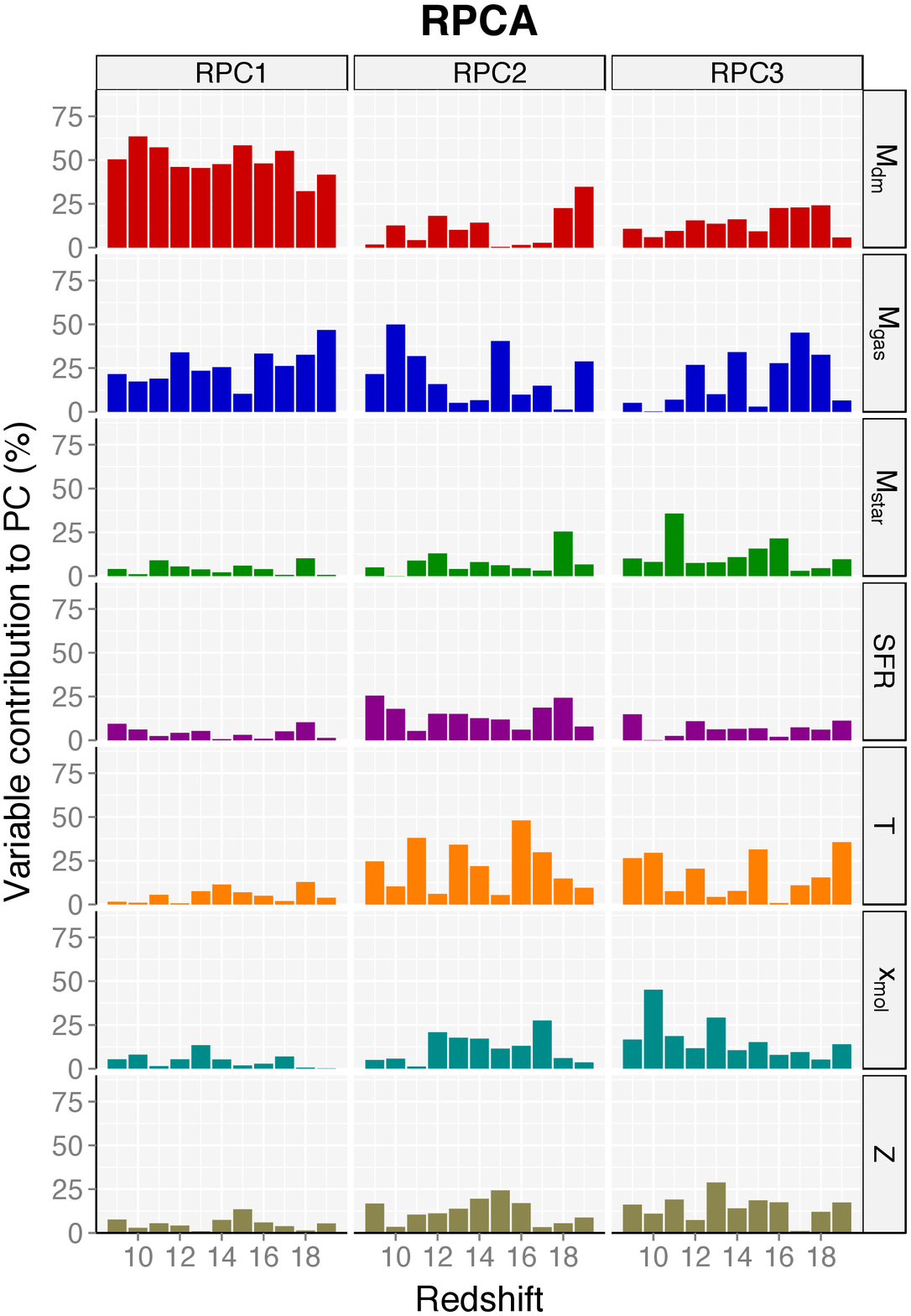}
\caption{Variable contribution to the first three components   as a function of redshift. From top to bottom: $M_{\rm dm}$ (red), $M_{\rm gas}$ (blue), $M_{\rm star}$ (green), \textit{SFR} (purple), $T$ (orange), $\mathrm{x_{mol}}$ (cyan), $Z$ (khaki). Left panel,  PCA; right panel, RPCA}
\label{fig:PC_bar}
\end{minipage}
\end{figure*}

Fig.   \ref{fig:PC_bar} shows the relative contribution of each parameter to the first three PCs (RPCs)  on the left (right) side.    For the PCA case,   $M_{\mathrm{dm}}$ and $M_{\mathrm{gas}}$ dominate PC1 at $z > 14$ (no less than $\sim$ 62 per cent), followed by a smaller contribution of SFR and  $\mathrm{x_{mol}}$. Nevertheless, as gas collapses into potential wells, the relative contribution from $M_{\mathrm{gas}}$ increases,  surpassing $M_{\mathrm{dm}}$ at $z \approx 15$. The dominant contribution of $Z$ and $\mathrm{x_{mol}}$ to PC1 at $z \approx 14$ indicates a critical epoch for the cosmic chemical enrichment (see also discussion above), triggered by a rapid variation of $\mathrm{x_{mol}}$, followed by a wide metal pollution at $z \approx 13$. After a decline in the chemical enrichment rate, a second peak in $Z$  occurs at $z \approx 10$. This self-regulated, oscillatory behavior is caused by the simultaneous coexistence of cold pristine-gas inflows and hot metal enriched outflows that create hydro instabilities and turbulent patterns with Reynolds numbers $\sim 10^8-10^{10}$ \citep[see e.g. Fig. 2 from][]{maio2011a}.
Finally at $z = 9$, $M_{\mathrm{dm}}$ and $M_{\mathrm{gas}}$ have become almost subdominant, since PC1 is mainly led by $T$ and $Z$, as a result of the ongoing cosmic heating from SF and thermal feedback. The dominance by $T$ to PC1 at this redshift  occurs due to the presence of  some small (see Fig. \ref{fig:variables_evol}), high-temperature objects,
whose properties are contaminated by hot enriched material
at $T \gtrsim10^5$ K.

An inspection of PC2  reveals the \textit{supporting roles} during  the galaxy formation process. The PC1 peak in $Z$ at redshift 13 is preceded by a strong contribution of \textit{SFR} and halo masses to PC2, 
while the second PC1 peak in $Z$, around $z \simeq 10$, is anticipated by an increasing contribution to PC2 from the formed stars, which later explode as SNe and start the metal enrichment of  the Universe. The first rise of PC2 at $z \gtrsim 14 $, dominated by SFR,   occurs because   the protogalaxies at this epoch are experiencing the  first bursts of SF.  Nevertheless,    not all of them have necessarily formed stars already. Whilst  the second peak is composed of a more balanced contribution from SFR and $M_{\rm star}$. The oscillatory   behavior might be caused by the  competitive effects of different feedback
mechanisms: the gas undergoing  SF
is heated by SN explosions and it  is  inhibited to continuously  form
stars (mostly in smaller structures that suffer significantly
gas evaporation processes); while shock compressions and spreading of metals in the
medium enhance gas cooling capabilities and  consequently induce more
SF.  The former preferentially occurs in bigger objects
that can keep  and re-process their metals because of the deeper potential wells. 

PC3  is nearly  negligible in the whole redshift range  aside $z = 14$,   where $ \mathrm{x_{mol}}$ dominates  the general behavior. This epoch is   preceded by a significant contribution from $M_{\rm star}$  at $z = 15$.  A comparison with Fig. $\ref{fig:variables_evol}$ reveals that this behavior coincides with a growth  in the $\mathrm{x_{mol}}$ variance at the same redshift. 
This indicates a  transition in the    regular trend of increasing $\mathrm{x_{mol}}$
with increasing mass at  $z \sim 15-16$, when initial
collapse phases boost $\mathrm{x_{mol}}$  up to  $10^{-3}$. This rapid growth of $\mathrm{x_{mol}}$ preferentially occurs in galaxies of
$\sim 10^5-10^6 M_{\bigodot}$, that are forming their first stars and have not been previously affected by   feedback mechanisms. At $z \lesssim15$, feedback
effects from Pop III  forming galaxies become responsible
for increasing the variance of  $\mathrm{x_{mol}}$ by  several  
orders of magnitude, either by dissociating molecules, or by partially enhancing their formation by shocks
and gas compression \citep[e.g., ][]{Ricotti2001,Whalen2008,pm2012}.

Looking the RPCA, the RPC1 is dominated by halo masses during all cosmic evolution (no less than 68 per cent), with other baryonic properties  relegated to RPCs of higher orders. 
Some caution is needed to interpret these results. The higher level of compressibility presented by  RPCA is  a  direct consequence of attributing a smaller   weight to rare events. 
Therefore, if one intends   to describe all  haloes properties using the fewest parameters possible, RPCA appears to succeed, since it states that as a first approximation,  the total halo mass  is  the main factor to describe all  other  properties.  The mass  determines the potential well and consequently the ability of the halo to form stars, retain  the metals, etc, therefore  roughly dictating  the baryonic dynamics at a first sight. 
Since RPCA ascribes   a lower weight  to the tails of  each parameter distribution, the physical interpretation may become less evident for the highest RPCs. 	
However,  we can still see the importance of $Z$, $\mathrm{x_{mol}}$ and SFR, with the difference that now they are considered second order effects,  hence  starting to be  dominant from the  RPC2 forward.  To better understand these differences between RPCA and PCA we discuss the strength  with   which each variable is related to one another as follows.

\subsection*{MIC}

\begin{figure*}
\label{fig:scatter}
\begin{minipage}[b]{1\linewidth}
\centering
\includegraphics[width=1.025\columnwidth]{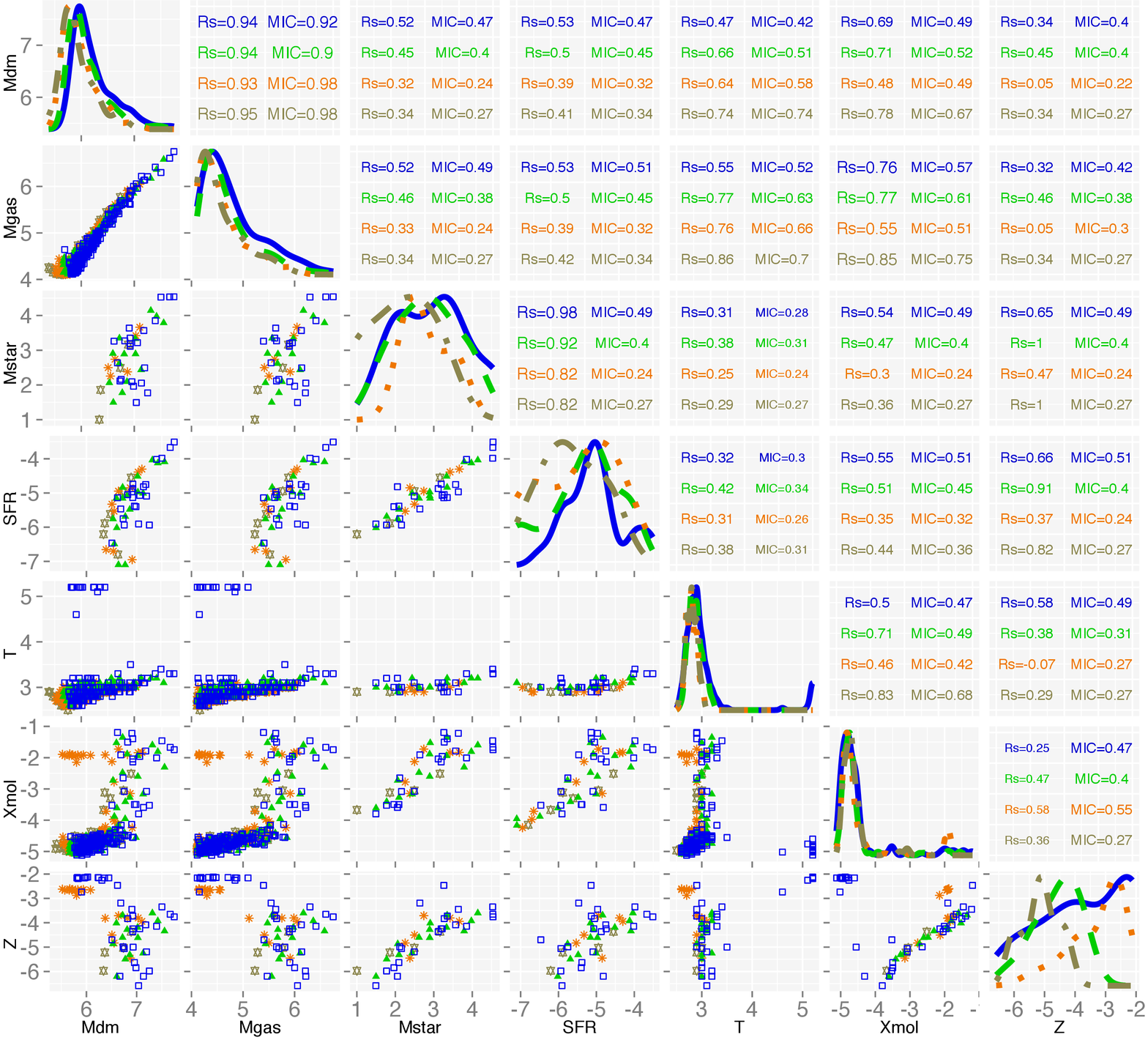}
\caption{Correlations between different halo properties at redshifts \emph{z} = 9 (blue solid lines and squares),  \emph{z} = 11 (green dashed  lines and triangles); \emph{z} =13 (orange dotted lines  and asterisks),     \emph{z} = 15 (khaki dot-dashed  lines and stars).
 The panels on the diagonal show the 
density distributions of the seven  parameter.   The bottom half matrix shows a scatter plot for each pair-variable combination and the top half matrix shows the MIC and Spearman rank coefficients  for each redshift. To guide the eyes,  the values  are colored by redshift and the font  size  is  proportional to the strength of the correlation. While the coefficients were estimated in the original parameters, the figures show the variables transformed by $\log{(M_{\rm dm}/M_{\bigodot})}$, $\log{(M_{\rm gas}/M_{\bigodot})}$, $\log{(M_{\rm star}/M_{\bigodot})}$, $\log{(SFR/M_{\bigodot}\rm yr^{-1})}$, $\log{(T/K)}$, $\log{(\mathrm{x_{mol}})}$ and $\log{(Z/Z_{\bigodot})}$,     for better visualization. }

\end{minipage}
\hspace{0.075cm}
\end{figure*}

Fig.  \ref{fig:scatter} shows how the seven  halo properties correlate to each other.  
The main diagonal of Fig.  \ref{fig:scatter} shows the density  distribution of each variable at different redshifts\footnote{Highest redshifts are not shown, because the few number of haloes make the PDF estimate meaningless.} (a zoomed  version of half-violin presented  in Fig. \ref{fig:variables_evol}). The majority of the parameters have a well behaved distribution, with small variations in  its shape   during the cosmic evolution,  while quantities related to  the stellar feedback ($M_{\rm star}, \textit{SFR}, Z$) have their distribution shaped during the  transition from a regime without SF  activity  at $z \gtrsim 16$   to the burst of  SFR around  $z \lesssim15$. 
The lower triangular part of the panel shows scatter plots for each variable combination colored accordingly to  their redshift.

\begin{figure}
\begin{minipage}[b]{1\linewidth}
\centering
\includegraphics[width=1.075\columnwidth]{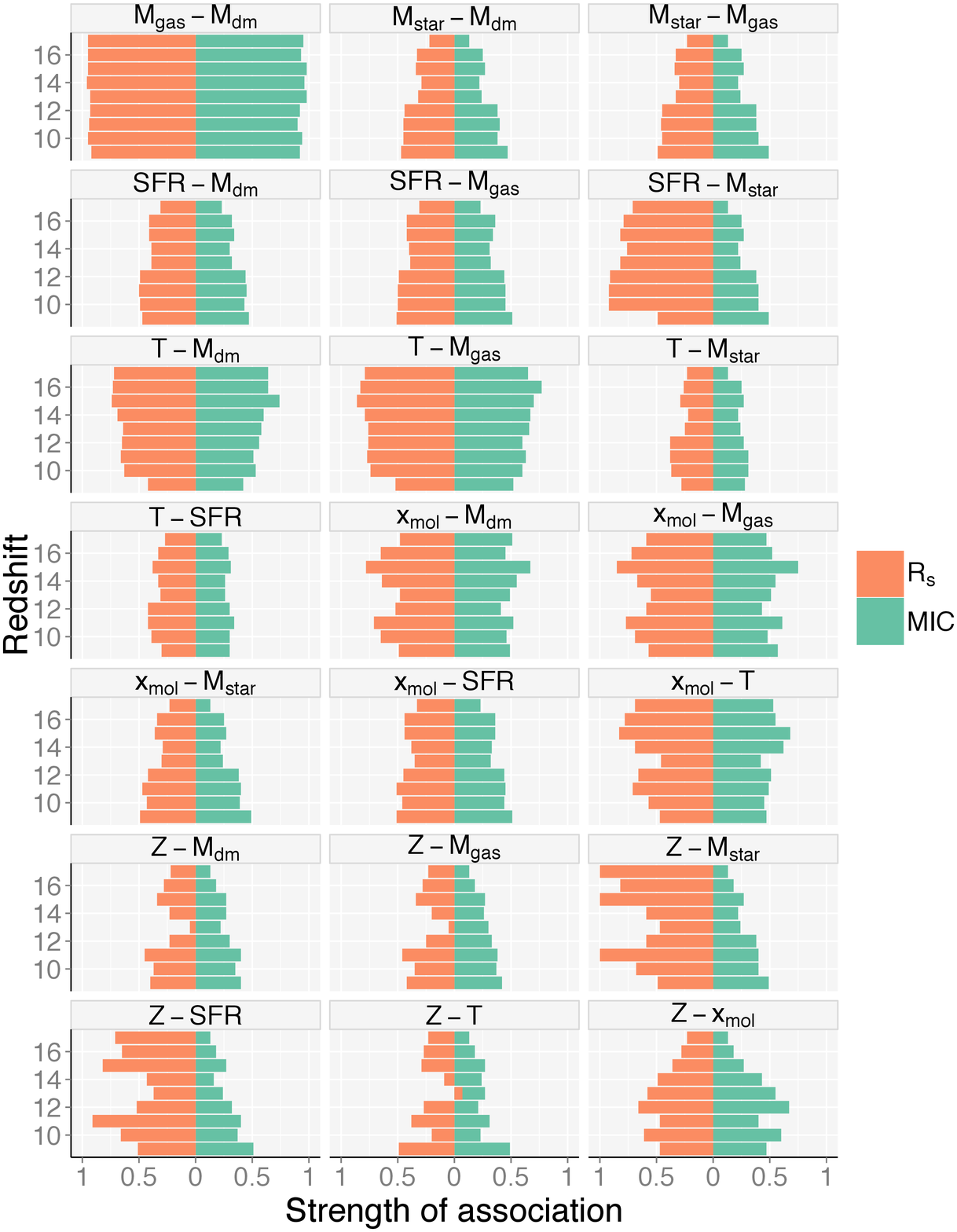}
\caption{Each panel shows MIC and Spearman correlations between different halo properties.  $R_{s}$   correlation  is shown on the   left scale  with orange bars, while the   MIC is represented by green bars on the right. }
\label{fig:MINE2}
\end{minipage}
\hspace{0.075cm}
\end{figure}

Fig.  \ref{fig:MINE2} shows  MIC and $R_s$ for each combination of parameters as a function of  redshift\footnote{We do not present results for $z>17$, because of the high number of zeros in the matrix makes the  correlation  measurements  unreliable.}.  At high redshift,  due to the poor statistics (less than 30 haloes at $z = 19$,  with a considerably amount of  null parameters), most variables are uncorrelated, receiving a low score by both $R_s$ and MIC. As expected $M_{\mathrm{gas}}$,  $M_{\mathrm{dm}}$ and $T$ are strongly correlated,  receiving the  highest  values.  This is consistent with the fact that PC1 dominates at $z > 16$ and is basically dictated by $M_{\rm dm}$ and $M_{\rm gas}$. The result suggests that at higher redshifts,  haloes are much simpler objects and their properties are   basically controlled  by their masses. Comparing with Fig. \ref{fig:PC_bar}, it seems that the correlation between halo mass and $T$ shows a better agreement with RPCA, which makes of   $T$ a factor almost as important as  $M_{\mathrm{gas}}$ and  $M_{\mathrm{dm}}$ in the determination of RPC1.
\\
The  molecular content, which  is directly dependent on the local gas density and \textit{T}, shows a correlation with   $Z$ that   increases at lower redshifts until $z \approx 12$. This  trend is in agreement with the dominance of  $\mathrm{x_{mol}}$ and $Z$ on  PC1 and RPC2  at $z \approx 13-14$,  caused by   the increase in the contribution of the \textit{SFR} to PC2 and RPC2   at  earlier redshifts. 
\\
At $z \gtrsim13-14$, $\mathrm{x_{mol}}$ keeps a regular trend of increasing with halo mass. Nevertheless,  the \textit{SF} activity at $ z \lesssim 13$ leads to a dispersion of $\mathrm{x_{mol}}$   followed by a metal enrichment process,   as discussed in Section \ref{sec_results}. Also $M_{\mathrm{gas}}$ shows a stronger correlation with  $\mathrm{x_{mol}}$ than with other quantities like SFR and $Z$, which indicates the crucial role of $\mathrm{x_{mol}}$ to initiate SF and consequent metal pollution from Pop  III and Pop II/I regimes in primordial galaxies. Comparing with Fig. \ref{fig:PC_bar}, we see that RPCA  better apprehends  this effect. At high redshift,  with the exception of  $z = 16$, where the peak in RPC2  is caused by the first stages of metal enrichment (Fig. \ref{fig:variables_evol}), $\mathrm{x_{mol}}$  maintains  a dominant  contribution to RPC2,   together with halo mass. 
 \\ 
The correlation between SFR with  $M_{\rm gas}$ and $M_{\rm dm}$ is roughly linear,  increasing  at later times. 
This may be explained by the wider  spread  of SFR  in low massive  haloes  at $z \gtrsim 14$, which is caused  by  gas evaporation processes due to SN explosions, in contrast with later structures that have a more sustained SF activity.  Albeit both PCA and RPCA are sensitive to this effect, RPCA  ascribes
 a lower weight to the SFR than to $\mathrm{x_{mol}}$,  in accordance to the correlation analysis.

 A surprising    disagreement between MIC and $R_{s}$  appears when comparing $Z$, $M_{\mathrm{star}}$ and  SFR.  $R_s$ suggests a nearly  perfect correlation between $Z$ and $M_{\mathrm{star}}$, while MIC found no significant association at the  highest redshifts.
This highlights the robustness of MIC with skewed and sparse data.  In this redshift range, $z \gtrsim14$,  there are very few haloes with non-null $Z$ and $M_{\mathrm{star}}$ values (Fig. \ref{fig:variables_evol}). Therefore, the high $R_s$ score  for these two quantities is misleading, as confirmed by a visual inspection of their corresponding distributions (Figs.  \ref{fig:variables_evol} and \ref{fig:scatter}). The same argument holds for the comparison between $Z$-SFR, and $M_{\mathrm{star}}$-SFR. 
During the course of cosmic evolution though, the correlations between the properties of the haloes tighten and both $R_s$ and MIC converge for most of them at $z = 10$ (with $R_s$ slightly overestimating the strength of correlation compared to MIC), as shown in Fig.  \ref{fig:MINE2}.

\section{Conclusions }
\label{sec_conclusions}

We investigate the redshift evolution of the gas properties of primordial galaxies using RPCA and MIC statistics making a comprehensive comparison with  standard approaches.

This is the first attempt to probe the baryon properties of early mini-haloes  and the effects of feedback processes by means of a highly  solid statistical approach.
We explore the correlation of different baryonic properties as expected from numerical $N$-body, hydrodynamical, chemistry simulations including gas molecular and atomic cooling, SF, stellar evolution, metal spreading and feedback effects.

The wide  range of redshifts analyzed here  ($9 \lesssim z \lesssim 19$) allowed  us to perform an unprecedented  study of the temporal evolution of the PC contribution  to the total variance of the halo properties.
The standard PCA needs two  PCs to  explain more than 92 per cent of the data variance (in the greater  part of redshifts studied here)  with  PC1 dropping below 50 per cent at lower redshifts. The first  RPC from RPCA analysis explains no less than 84 per cent of all data variance anytime, with two first RPCs explaining more than 95 per cent of the total robust variance.  

First SF episodes and feedback mechanisms cause a drop of PC1 at $z \sim 14$, when a sharp variation in the PCs behavior marks the onset of cosmic metal enrichment. At $z > 14$ the halo properties are basically dictated by the halo mass.    Among the advantages in using RPCA is the possibility to increase the capability to reduce the dimensionality of the original dataset, although at the cost to be less sensitive to rare events that may be physically relevant.  Since  RPCA  ranks  the contribution of variables to the RPCs in  better agreement with   their levels of correlation.  It  seems to be in better agreement with our  independent  MIC and $R_{s}$ correlation analysis. 	

An inspection in the first and second PCs reveals some interesting facts.  The PC1 peak in $Z$ at redshift 13 is preceded by a strong contribution of SFR and halo masses to PC2.
While the second PC1 peak in $Z$, around $z \simeq 10$, is anticipated by an increasing contribution to PC2 by the formed stars, which later explode as SNe and enrich the Universe. This indicates the importance of stellar evolution in shaping   baryon properties in primordial haloes. A similar   trend holds for RPCA although attenuated by the smoothing effect created by the use of robust statistics.  
  It is important to note,  however,  that the relatively small number of  haloes studied here might   lessen the robustness  of our results at very high redshifts. Therefore,  future investigations  of similar  techniques into larger simulations boxes  is  highly recommended. 

 Overall  $R_s$ agrees reasonably with MIC, but MIC seems to be more robust to study highly sparse data regimes (like  at early epochs).
All gas properties, aside $M_{\mathrm{gas}}$, $M_{\mathrm{dm}}$ and $T$, are weakly correlated at high redshift. Nevertheless, due to the interplay between chemical and mechanical feedback from the ongoing stellar formation and the consequent back reaction on the thermal behavior of the surrounding medium, baryonic quantities start to present a moderate to high level of correlation as redshift decreases.
In particular, $\mathrm{x_{mol}}$ shows the highest level of correlation with $M_{\mathrm{gas}}$, followed by $T$,  SFR, $M_{\rm star}$ and $Z$ respectively.
In general, structure formation processes depend not only on the dark matter halo properties, but also on the local thermodynamical state of the gas, which is, in turn, affected by cooling, \textit{SF} and feedback.   Our analysis suggests that all the gaseous properties have a stronger correlation with $M_{\mathrm{gas}}$ than with $M_{\mathrm{dm}}$, while $M_{\mathrm{gas}}$ has a deeper correlation with ${\mathrm{x_{mol}}}$ than with \textit{Z} or SFR.
The relevance of  the molecular content for the baryon properties represents the physical origin of gas collapse and concentration,  crucial to initiate SF and consequent metal pollution from Pop  III and Pop  II/I regimes in primordial galaxies.
\\ 
This work represents a  leap forward   in the statistical analysis of $N$-body/hydro simulations, performed by means of RPCA and MIC into a cosmological context. We therefore stress that the use of dimensionality reduction algorithms and mutual information based techniques in numerical simulations might be a precious instrument for future investigations, thanks to their  potential to unveil non-trivial relationships, which may go undetected by standard methods.


\section*{Acknowledgements}
U.M. has received funding from the European Union Seventh Framework Programme (FP7/2007-2013) under grant agreement n.267251.
For the bibliographic research we made use of the NASA Astrophysics Data System archive. We thank the referee by the very  useful comments, helping to enhance the article presentation. 
\footnotesize{

}

\bsp

\label{lastpage}
\end{document}